\documentclass[%
superscriptaddress,
preprint,
 amsmath,amssymb,
 aps,
floatfix,
]{revtex4-1}

\usepackage{graphicx}
\usepackage{dcolumn}
\usepackage{bm}
\usepackage[mathlines]{lineno}

\begin{document}

\title{A CMOS silicon spin qubit}

\author{R. Maurand}
 \email{romain.maurand@cea.fr}
\author{X. Jehl}
\author{D. Kotekar Patil}
\author{A. Corna}
\author{H. Bohuslavskyi}
\affiliation{University Grenoble Alpes, F-38000 Grenoble, France}
\affiliation{CEA, INAC-PHELIQS, F-38000 Grenoble, France}
\author{R. Lavi\'eville}
\author{L. Hutin}
\author{S. Barraud}
\author{M. Vinet}
\affiliation{University Grenoble Alpes, F-38000 Grenoble, France}
\affiliation{CEA, LETI, MINATEC Campus, 17 rue des Martyrs, F-38054 Grenoble, France}
\author{M. Sanquer}
\author{S. De Franceschi}
 \email{silvano.defranceschi@cea.fr}
\affiliation{University Grenoble Alpes, F-38000 Grenoble, France}
\affiliation{CEA, INAC-PHELIQS, F-38000 Grenoble, France}

\maketitle

\textbf{Silicon, the main constituent of microprocessor chips, is emerging as a promising material\cite{Maune2012,Pla2012,Kawakami2014,Veldhorst2014,Muhonen2014,Veldhorst2015} for the realization of future quantum processors\cite{Kane1998,Loss1998}. Leveraging its well-established complementary metal-oxide-semiconductor (CMOS) technology would be a clear asset to the development of scalable quantum computing architectures\cite{Hill2015,Pica2016} and to their co-integration with classical control hardware\cite{Levy2011}.  Here we report a silicon quantum bit (qubit) device made with an industry-standard fabrication process\cite{Barraud2012}. The device consists of a two-gate, p-type transistor with an undoped channel. At low temperature, the first gate defines a quantum dot (QD) encoding a hole spin qubit, the second one a QD used for the qubit readout. All electrical, two-axis control of the spin qubit is achieved by applying a phase-tunable microwave modulation to the first gate. Our result opens a viable path to qubit up-scaling through a readily exploitable CMOS platform.}

Localized spins in semiconductors can be used to encode elementary bits of quantum information~\cite{Kane1998,Loss1998}. Spin qubits were demonstrated in a variety of semiconductors, starting from GaAs-based heterostructures \cite{Petta2005,Koppens2006, Hanson2006}. In this material, and all III-V compounds in general, electron spins couple to the nuclear spins of the host crystal via the hyperfine interaction resulting, in a relatively short inhomogeneous dephasing time, $\rm{T_2^*}$ (a few tens of ns in GaAs~\cite{Koppens2008}). This problem can be cured to a large extent by means of echo-type spin manipulation sequences and notch filtering techniques~\cite{Bluhm2010_2,Lange2010,Malinowski2016}. 
In natural silicon, however, the hyperfine interaction is weaker, being due to the $\approx 4.7\%$ content of $^{29}$Si, the only stable isotope with a non-zero nuclear spin. Measured $\rm{T_2^*}$ values range between $50$\,ns and $2$\,$\mu$s  \cite{Pla2012,Maune2012,Wu2014,Kawakami2014}. Experiments carried out on electron spin qubits in isotopically purified silicon ($99.99\%$ of spinless $^{28}$Si), have even allowed extending  $\rm{T_2^*}$ to $120$\,$\mu$s~\cite{Veldhorst2014}.
Following these improvements in spin coherence time, silicon-based spin qubits classify among the best solid-state qubits, at the single qubit level. Recently, the first two-qubit logic gate with control-NOT functionality was also demonstrated~\cite{Veldhorst2015}, marking the next essential milestone towards scalable processors.

Surface-code quantum computing architectures, possibly the only viable option to date, require large numbers (eventually millions) of qubits individually controlled with tunable nearest-neighbor couplings~\cite{Bravyi2008,Dennis2002}. Their implementation is a considerable challenge since it implies dealing with issues such as device-to-device variability, multi-layer electrical wiring, and, most likely, on-chip classical electronics (amplifiers, multiplexers, etc) for qubit control and readout. This is where the well-established CMOS technology becomes a compelling tool. A possible strategy is to export qubit device implementations developed within academic-scale laboratories into large-scale CMOS platforms. This approach is likely to require significant process integration development at the CMOS foundry. Here we present an alternative route, where an existing process flow for the fabrication of CMOS transistors is taken as a starting point, and it is adapted to obtain devices with qubit functionality.    

\begin{figure}
\includegraphics[width=0.9\columnwidth]{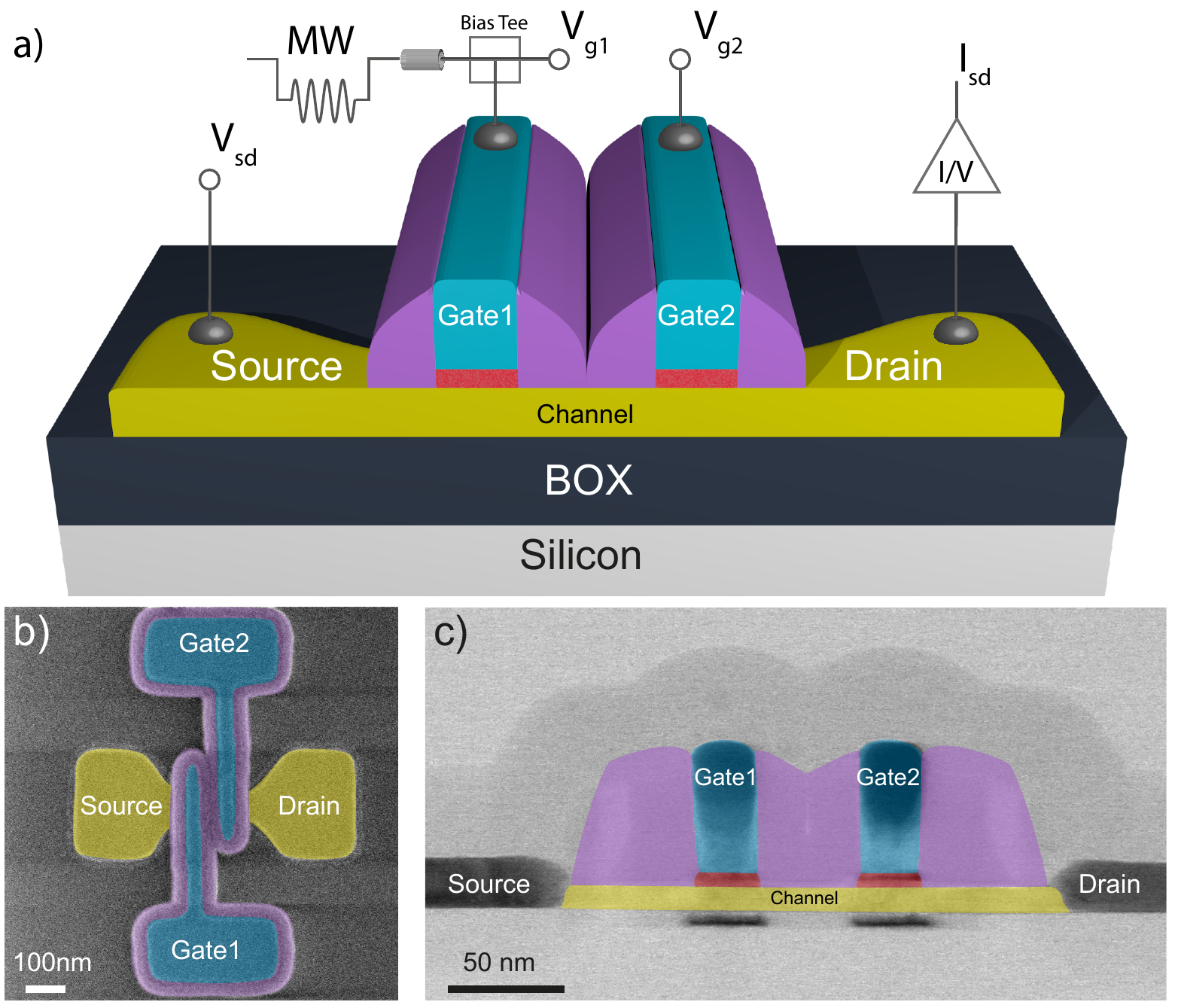}
\caption{\textbf{CMOS qubit device. a}, Simplified 3-dimensional schematic of a SOI nanowire field-effect transistor with two gates Gate\,1 and Gate\,2. Using a bias-T, Gate 1 is connected to a low-pass-filtered line, used to apply a static gate voltage $V_{g1}$, and to a 20-GHz bandwidth line, used to apply the high-frequency modulation necessary for qubit initialization, manipulation and readout. \textbf{b}, Colorized device top view obtained by scanning electron microscopy just after the fabrication of gates and spacers. \textbf{c}, Colorized transmission-electron-microscopy image of the device along a longitudinal cross-sectional plane.} 
\label{fig1} 
\end{figure}

We use a microelectronics technology based on $300$ mm Silicon-On-Insulator (SOI) wafers. Our qubit device, schematically shown in Fig.~\ref{fig1}a, is derived from silicon nanowire field-effect transistors~\cite{Barraud2012}. It consists of a $10$-nm-thick and $20$-nm-wide undoped silicon channel with p-doped source and drain contact regions, and two $\approx30$\,nm wide parallel top gates, side covered by insulating silicon nitride spacers. A scanning-electron-microscopy top view, and a transmission-electron-microscopy cross-sectional view are shown in Fig.~\ref{fig1}b and 1c, respectively.
At low temperature, hole QDs are created by charge accumulation below the gates \cite{Voisin2016}. The double gate layout enables the formation of two QDs in series, QD1 and QD2, with occupancies controlled by voltages $V_{g1}$ and $V_{g2}$ applied to gates 1 and 2, respectively (see supplementary section 2). We tune charge accumulation to relatively small numbers, N, of confined holes ($N\approx 10$ per dot). In this regime, the QDs exhibit a discrete energy spectrum with level spacing $\delta E$  in the 0.1 - 1\,meV range, and Coulomb charging energy $U \approx 1 - 10$\,meV.

In a simple scenario where spin-degenerate QD levels get progressively filled by pairs of holes, each QD carries a spin $S=1/2$ for N=odd and a spin $S=0$ for N=even. By setting N=odd in both dots two spin-1/2 qubits can be potentially encoded, one for each QD. This is equivalent to the (1,1) charge configuration, where the first and second digits denote the charge occupancies of QD1 and QD2, respectively. In practice, here we shall demonstrate full two-axis control of the first spin only, and use the second spin for initialization and readout purposes. Tuning the double QD to a parity-equivalent (1,1) $\to$ (0,2)  charge transition, initialization and readout of the qubit relies on the so-called Pauli spin blockade mechanism~\cite{Ono2002, Hanson2006}. In this particular charge transition, tunneling between dots can be blocked by spin selection rule. Basically, for a fixed, say ``up'', spin orientation in QD2, tunneling will be allowed if the spin in QD1 is ``down" and it will forbidden by the Pauli exclusion principle if the spin in QD1 is ``up" \textit{i.e.} a triplet (1,1) state is not coupled to the singlet (0,2) state. This charge/spin configuration can be identified through characteristic experimental signatures~\cite{Danon2009,Nadj-Perge2010,Li2015} associated with the Pauli blockade effect discussed above (see supplementary section 3).

\begin{figure}
\includegraphics[width=1\columnwidth]{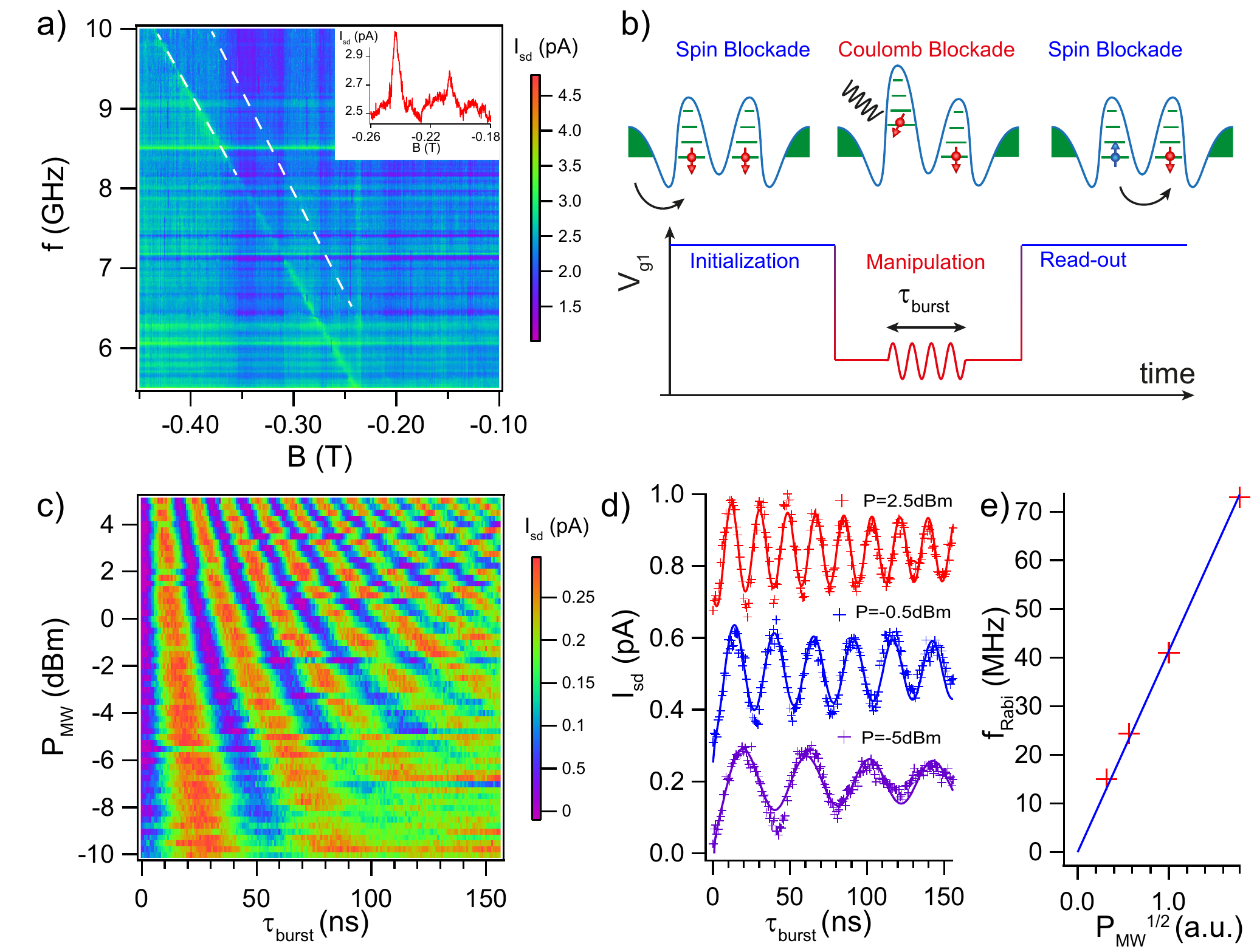}
\caption{\textbf{Electrically driven coherent spin manipulation. a}, Color plot of $I_{\rm sd}$ as a function of magnetic field, $B$, and MW frequency, $f$. Electrically driven hole spin resonance is revealed by two enhanced current ridges highlighted by white dashed lines. Inset: horizontal cut at $f=5.4$\,GHz. \textbf{b}, Schematic representation of the spin
manipulation cycle and corresponding gate-voltage ($V_{\rm g1}$) modulation pattern. \textbf{c}, Color plot of Rabi oscillations for a range of microwave power at $f=8.938$\,GHz and $B=0.144$\,mT. Current has been averaged for $1$\,s for each data point. \textbf{d}, Rabi oscillations for different power taken from \textbf{c} fitted~\cite{Koppens2007} with $A\cos(2\pi f_{\rm Rabi}\tau_{\rm burst}+\phi)/\tau_{\rm burst}^\alpha$. Rabi frequencies are $24$\,MHz, $39$\,MHz and $55$\,MHz for $P=-5$\,dBm, $P=-0.5$\,dBm and $P=2.5$\,dBm respectively. \textbf{e}, Rabi frequency \textit{versus} on microwave amplitude with a linear fit. } 
\label{fig2} 
\end{figure}

We now turn to the procedure for spin manipulation. In a recent work on similar devices with only one gate, we found that hole g-factors are anisotropic and gate dependent~\cite{Voisin2016}, denoting strong spin-orbit coupling (see also Ref. \cite{Li2015}). This implies the possibility to perform electric-dipole spin resonance (EDSR), namely to  drive coherent hole-spin rotations by means of microwave-frequency (MW) modulation of a gate voltage (see supplementary section 4). Here we apply the MW modulation to Gate\,1 in order to rotate the spin in QD1. Spin rotations result in the lifting of spin blockade.
In a measurement of source-drain current $I_{\rm sd}$ as a function of magnetic field, $B$, (perpendicular to the chip) and MW frequency, $f$, EDSR is revealed by narrow ridges of increased current\cite{Nadj-Perge2010}. The data set in Fig. \ref{fig2}a shows two of such current ridges, one clearly visible and the other one rather faint. Both ridges follow a linear $f(B)$ dependence consistent with the spin resonance condition $hf = g \mu_B B$, where $h$ is Planck's constant, $\mu_B$ the Bohr magneton, and $g$ the hole Land\'e g-factor along the magnetic-field direction. From the slopes of the two ridges we extract two g-factor values, $g_1 = 1.92$ and $g_2 = 1.63$ comparable to those reported before~\cite{Voisin2016}. Based on the relative intensities of the current ridges we ascribe these g-factor values to QD1 and QD2, respectively. We have observed similar EDSR features at other working points (i.e. different parity-equivalent (1,1) $\to$ (0,2) transitions) and in two distinct devices (see supplementary section 4).

\begin{figure}
\includegraphics[width=0.75\columnwidth]{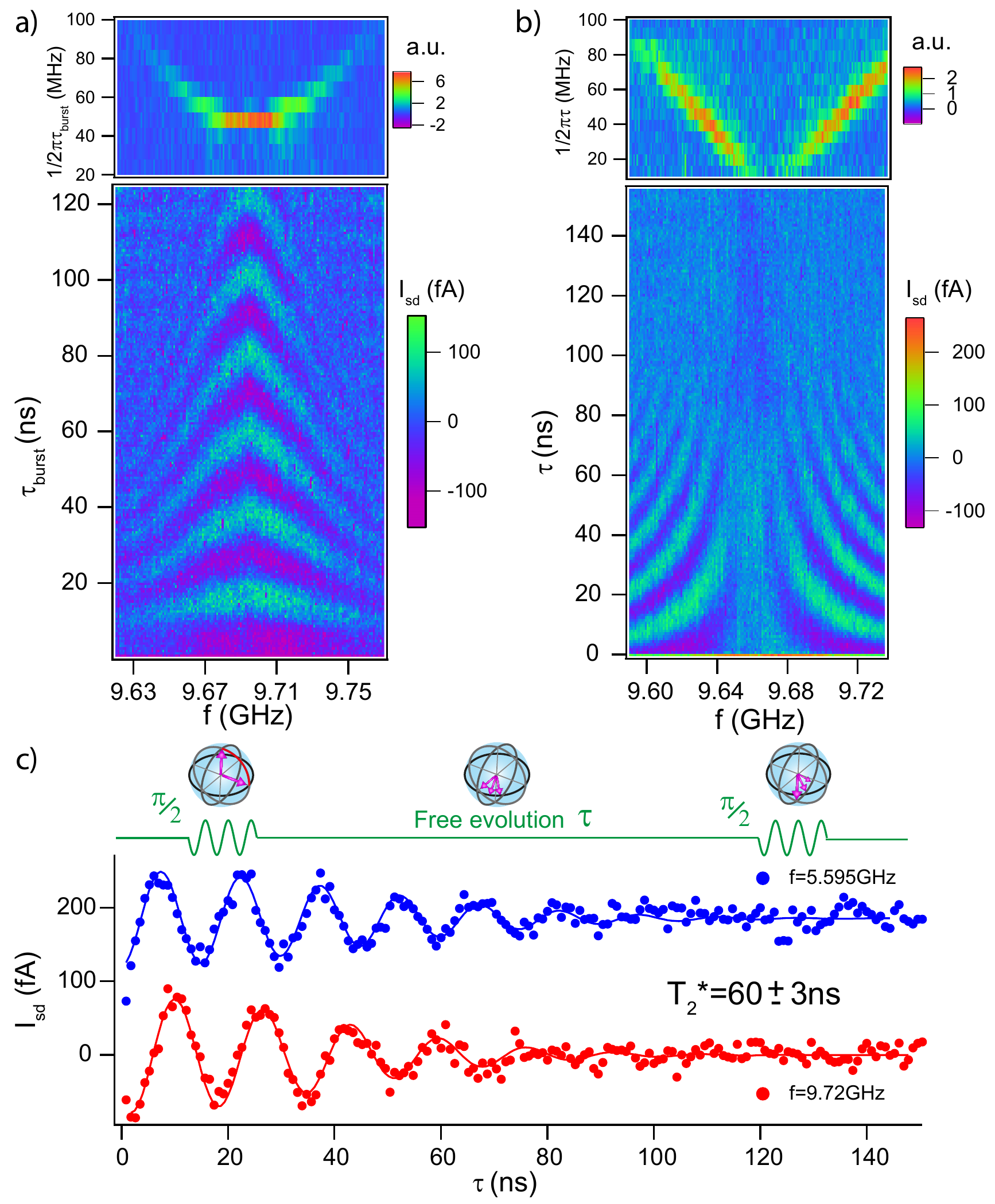}
\caption{\textbf{Frequency dependence of Rabi oscillations and Ramsey fringes. a}, Bottom panel: $I_{\rm sd}(f,\tau_{\rm burst})$ at $B=0.155$\,mT and $P_{\rm MW}=3$\,dBm. 
Every data point was averaged for $600$\,ms and, for each $f$, the average current was subtracted. 
Top panel:  Fourier transform of the data in the bottom panel showing the expected hyperbolic dependence of $f_{\rm Rabi}(f)$. 
\textbf{b}, Bottom panel:  $I_{\rm sd}(f,\tau)$, where $\tau$ is the waiting time between two 7-ns-long $\frac{\pi}{2}$ bursts. 
Every data point was obtained with a $2$\,s integration time and the average current was subtracted.
This data set, taken at $B=0.155$\,mT and $P_{\rm MW}=8$\,dBm, shows a characteristic Ramsey-interference pattern. 
%
%
Top panel: Fourier transform of the data in the bottom panel showing the expected linear evolution of the Ramsey fringes frequency depending on $f$. \textbf{c}, Ramsey sequence manipulation scheme (top), and two $I_{\rm sd}(\tau)$ data sets corresponding to vertical cuts in \textbf{b} for $f=5.595$\,GHz and $f=9.720$\,GHz  Solid lines are fits to $A\cos(\Delta f \tau +\phi)\exp(-(\tau/T_2^*)^2)$. The data in blue have an upward offset of $250$\,fA. } 
\label{fig3} 
\end{figure}

To perform controlled spin rotations, and hence demonstrate qubit functionality, we replace continuous-wave gate modulation with MW bursts of tunable duration, $\tau_{\rm burst}$. 
During spin manipulation, we prevent charge leakage due to tunneling from QD1 to QD2 by simultaneously detuning the double QD to a Coulomb-blockade regime \cite{Koppens2006}  (see Fig. \ref{fig2}b). 
Following each burst, $V_{\rm g1}$ is abruptly increased to bring the double dot back to the parity-equivalent (1,1) $\to$ (0,2) resonant transition. At this stage, a hole can tunnel from QD1 to QD2 with a probability proportional to the unblocked spin component in QD1 (i.e. the probability amplitude for spin-up if QD2 hosts a spin-down state). The resulting (0,2)-like charge state ``decays'' by emitting a hole into the drain, and a hole from the source is successively fed back to QD1, thereby restoring the initial (1,1)-like charge configuration. The net effect is the transfer of one hole from source to drain, which will eventually contribute to a measurable average current. (In principle, in case not all (1,1)-like states are Pauli blocked, the described charge cycle may occur more than once during the readout-initialization portion of the same period, until the parity-equivalent (1,1) $\to$ (0,2) becomes spin blocked again and the system is re-initialized for the next manipulation cycle.)

We chose a modulation period of $435$\,ns, of which $175$\,ns are devoted to spin manipulation and $260$\,ns to readout and initialization.  
Figure~\ref{fig2}c) shows $I_{\rm sd}$ as a function of MW power $P_{\rm MW}$ and $\tau_{\rm burst}$ at a spin-resonance condition for $B=144$\,mT. The observed current modulation is a hallmark of coherent Rabi oscillations of the spin in QD1, also explicitly shown by selected cuts at three different MW powers (Fig. ~\ref{fig2}d)). As expected, the Rabi frequency, $f_{\rm Rabi}$, increases linearly with the MW voltage amplitude, which is proportional to ${P_{\rm MW}}^{1/2}$ (Fig. ~\ref{fig2}e)). At the highest power, we reach a remarkably large $f_{\rm Rabi}\approx 85$\,MHz, comparable to the highest reported values for electrically controlled semiconductor spin qubits~\cite{VandenBerg2013}. 
Figure~\ref{fig3}a) shows a color plot of $I_{\rm sd}(f, \tau_{\rm burst})$ revealing the characteristic chevron pattern associated to Rabi oscillations~\cite{Kawakami2014}. The fast Fourier transform (FFT) of $I_{\rm sd}(\tau_{\rm burst})$, calculated for each $f$ value, is shown in the upper panel. It exhibits a peak at the Rabi frequency with the expected hyperbolic dependence on frequency detuning $\Delta f = f- f_0$, where $f_0 = 9.68$\,GHz is the resonance frequency at the corresponding $B=155$\,mT.

\begin{figure}
\includegraphics[width=0.8\columnwidth]{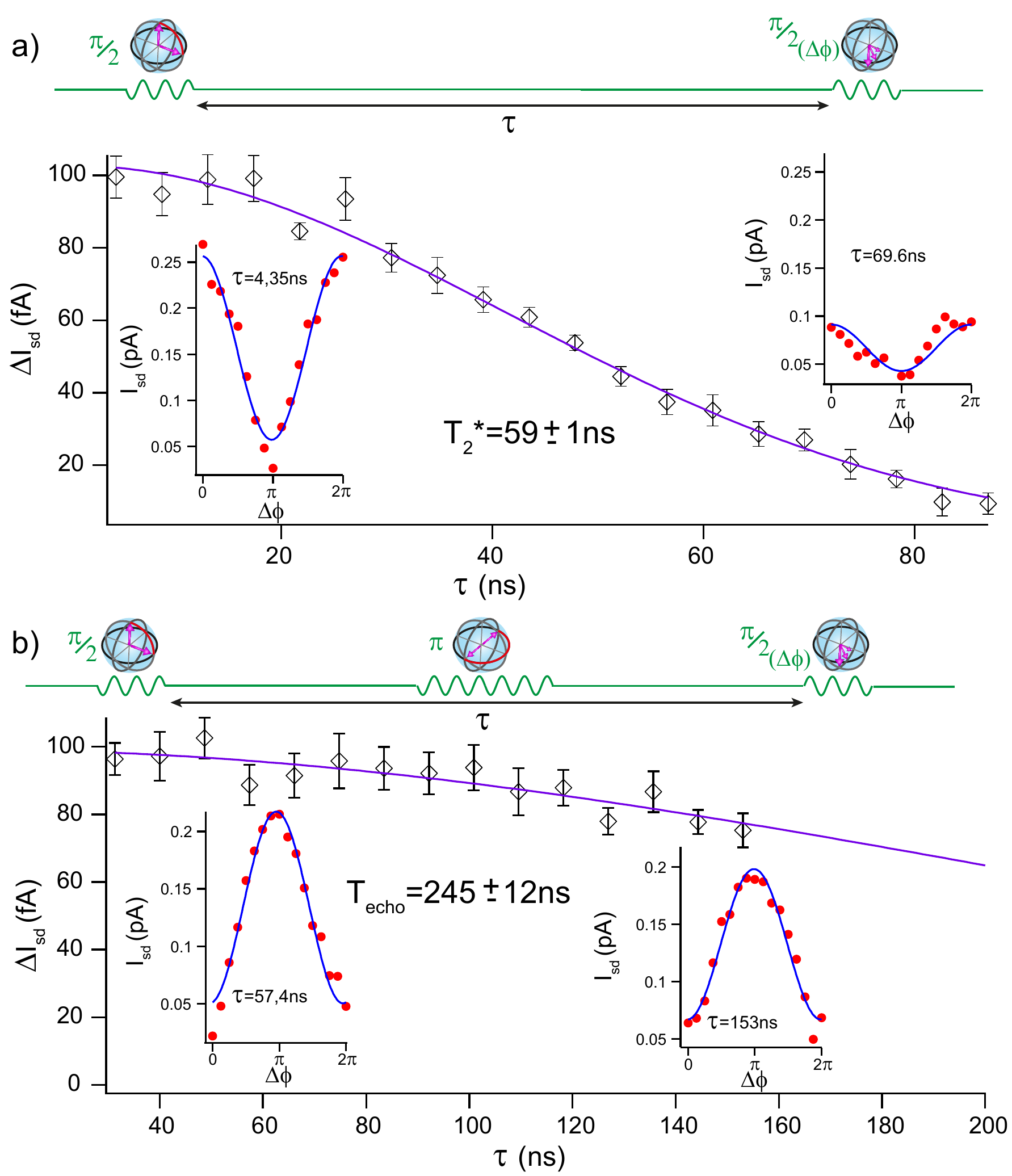}
\caption{\textbf{Two-axis qubit control and spin coherence times. a}, Amplitude $\Delta I_{\rm sd}$ of Ramsey oscillations \textit{vs} delay time $\tau$. For each $\tau$, the phase of the second $\pi/2$ pulse is shifted by  $\Delta \phi$ (see top diagram), which corresponds to a change in the rotation axis. Insets: full oscillations at short (4.35 ns) and long (69.6 ns) $\tau$ and corresponding sinusoidal fits enabling the extraction of $\Delta I_{\rm sd}$. The decay of $\Delta I_{\rm sd}(\tau)$ is fitted to $\exp(-(\tau/T_2^*)^2)$ giving $T_2^* = 59 \pm 1$ ns. \textbf{b}, Results of a Hahn-echo experiment, whose manipulation scheme is given in the top diagram. The duration of the refocusing $\pi$ pulse is $14$ ns. Insets:  full oscillations at relatively short (57.4 ns) and long (153 ns) $\tau$ and corresponding sinusoidal fits. 
The Hahn-echo oscillation amplitude $\Delta I_{\rm sd}$ decays on time scale longer than the largest $\tau$, which was limited to 160 ns to ensure a sufficiently fast repetition cycle, and hence a measurable readout current. 
Fitting $\Delta I_{\rm sd}(\tau)$ to $\exp(-(\tau/T_{\rm echo})^3)$ yields $T_{\rm echo} = 245 \pm 12$ ns.}
\label{fig4} 
\end{figure}

To evaluate the inhomogeneous dephasing time  $T_2^*$ during free-evolution, we perform a Ramsey fringes-like experiment, which consists in applying two short, phase coherent, MW pulses separated by a delay time $\tau$. The proportionality between the qubit rotation angle, $\theta$, and $\sqrt{P_{\rm MW}}\tau_{\rm burst}$ is used to calibrate both pulses to a $\theta=\frac{\pi}{2}$ rotation (see sketch in Fig.~\ref{fig3}c).
For each $f$ value, $I_{\rm sd}$ exhibits oscillations at frequency $\Delta f$ decaying on a timescale $T_2^* \approx 60$\,ns (see Fig. ~\ref{fig3}b)). Extracted current oscillations at fixed frequency are presented in Fig.~\ref{fig3}c). 
At resonance ($\Delta f=0$), the two pulses induce $\frac{\pi}{2}$ rotations around the same axis (say the x-axis of the rotating frame). The effect of a finite $\Delta f$ is to change the rotation axis of the second $\frac{\pi}{2}$ pulse relative to the first one. 
Alternatively, two-axis control can be achieved also at resonance ($\Delta f =0$) by varying the relative phase $\Delta\phi$ of the MW modulation between the two pulses. For a Ramsey sequence $\frac{\pi}{2}$-$\tau$-$\frac{\pi}{2}_{\Delta\phi}$, the first pulse induces a rotation around $x$ and the second one around $x$, $y$, $-x$ and $-y$ for $\Delta \phi=0$, $\frac{\pi}{2}$, $\pi$ and $\frac{3\pi}{2}$, respectively. The signal then oscillates with $\Delta \phi$ as shown in the insets to Fig.~\ref{fig4}a, and the oscillation amplitude vanishes with $\tau$ on a $T_2^*$ time scale (see Fig.~\ref{fig4}a). 

The intrinsic coherence time associated with the dynamics of the dominant dephasing source can be accessed by means of a Hahn-echo experiment, where a $\pi$ pulse is introduced half way between the two $\frac{\pi}{2}$ pulses, as sketched in Fig.~\ref{fig4}b. The amplitude of the oscillations in $\Delta\phi$ (insets to Fig.~\ref{fig4}b)) decays on a coherence time $T_{\rm echo}=245\pm12$\,ns. 
The relatively short $T_2^*$ and $T_{\rm echo}$ can hardly be explained by the dephasing from $\rm{Si}^{29}$ nuclear spins. In fact, even if little is known about the hyperfine interaction strength for confined holes in silicon, we would expect it to be even smaller than the one for electrons.~\cite{Testelin2009}. Alternative decoherence mechanisms could dominate, such as paramagnetic impurities, charge noise, or the stronger hyperfine interaction with boron dopants diffused from the contact regions. Further studies will be necessary to establish statistically relevant values for the coherence time scales and to identify their origin.

In essence, we have shown that a p-type silicon field-effect transistor fabricated within an industry-standard CMOS process line can exhibit hole spin qubit functionality with fast, all-electrical, two-axis control. In the prospect of realizing large-scale quantum computing architectures, this result opens a favorable scenario with some clear follow-up milestones. The next step is to advance from the simple, yet limited transistor-like structures studied here to more elaborate qubit designs, incorporating additional important elements such as single-shot qubit read-out, and enabling scalable qubit-to-qubit coupling schemes. In addition, a systematic investigation of qubit performances, including the benchmarking of hole qubits against their electron counterparts, has to be performed in the short term. The use of state-of-the-art CMOS technology, with its well-established fabrication processes and integration capabilities, is going to be a clear asset in all these tasks. At a later stage, it should also favor the co-integration of classical cryogenic control hardware.

\section*{Methods}
A detailed description of the device fabrication process is given in Supplementary section1.  
All measurements were performed in a dilution refrigerator with a base temperature of $T=10$\,mK. The direct source-drain current providing qubit readout was measured by means of a current/voltage amplifier with a gain of $10^9$. All low-frequency lines are low-pass filtered at base temperature with two stage RC filters. High frequency signals applied to Gate 1 come for a 20\,GHz bandwidth coaxial line with distributed 36\,dBm attenuation along the dilution fridge for thermalization. A home-made bias tee on the sample board allows combination of microwave and low-frequency signals on the gate. One channel of an arbitrary wave generator (Tektronix AWG5014C) is used to generate the two-level $V_{\rm g1}$ modulation driving the device between Coulomb blockade (qubit manipulation phase) and Pauli blockade (qubit readout and initialization). Two other channels of the AWG define square pulses to control the I and Q inputs of the MW source.  MW bursts and the two-level gate modulation are combined by means of a diplexer before reaching the dilution fridge.

\begin{acknowledgments}
We thank D. Est\`eve, M. Hofheinz, F. Kuemmeth, T. Meunier, J. Renard, N. Roch and D. Vion for their help as well as G. Audoit and C. Guedj for the TEM sample preparation and imaging. The research leading to these results has been supported by the European Union's through the research grants No. 323841, No. 610637, and No. 688539, as well as through the ERC grant No. 280043. 
\end{acknowledgments}


\begin{thebibliography}{30}%
\makeatletter
\providecommand \@ifxundefined [1]{%
 \@ifx{#1\undefined}
}%
\providecommand \@ifnum [1]{%
 \ifnum #1\expandafter \@firstoftwo
 \else \expandafter \@secondoftwo
 \fi
}%
\providecommand \@ifx [1]{%
 \ifx #1\expandafter \@firstoftwo
 \else \expandafter \@secondoftwo
 \fi
}%
\providecommand \natexlab [1]{#1}%
\providecommand \enquote  [1]{``#1''}%
\providecommand \bibnamefont  [1]{#1}%
\providecommand \bibfnamefont [1]{#1}%
\providecommand \citenamefont [1]{#1}%
\providecommand \href@noop [0]{\@secondoftwo}%
\providecommand \href [0]{\begingroup \@sanitize@url \@href}%
\providecommand \@href[1]{\@@startlink{#1}\@@href}%
\providecommand \@@href[1]{\endgroup#1\@@endlink}%
\providecommand \@sanitize@url [0]{\catcode `\\12\catcode `\$12\catcode
  `\&12\catcode `\#12\catcode `\^12\catcode `\_12\catcode `\%12\relax}%
\providecommand \@@startlink[1]{}%
\providecommand \@@endlink[0]{}%
\providecommand \url  [0]{\begingroup\@sanitize@url \@url }%
\providecommand \@url [1]{\endgroup\@href {#1}{\urlprefix }}%
\providecommand \urlprefix  [0]{URL }%
\providecommand \Eprint [0]{\href }%
\providecommand \doibase [0]{http://dx.doi.org/}%
\providecommand \selectlanguage [0]{\@gobble}%
\providecommand \bibinfo  [0]{\@secondoftwo}%
\providecommand \bibfield  [0]{\@secondoftwo}%
\providecommand \translation [1]{[#1]}%
\providecommand \BibitemOpen [0]{}%
\providecommand \bibitemStop [0]{}%
\providecommand \bibitemNoStop [0]{.\EOS\space}%
\providecommand \EOS [0]{\spacefactor3000\relax}%
\providecommand \BibitemShut  [1]{\csname bibitem#1\endcsname}%
\let\auto@bib@innerbib\@empty
\bibitem [{\citenamefont {Maune}\ \emph {et~al.}(2012)\citenamefont {Maune},
  \citenamefont {Borselli}, \citenamefont {Huang}, \citenamefont {Ladd},
  \citenamefont {Deelman}, \citenamefont {Holabird}, \citenamefont {Kiselev},
  \citenamefont {Alvarado-Rodriguez}, \citenamefont {Ross}, \citenamefont
  {Schmitz}, \citenamefont {Sokolich}, \citenamefont {Watson}, \citenamefont
  {Gyure},\ and\ \citenamefont {Hunter}}]{Maune2012}%
  \BibitemOpen
  \bibfield  {author} {\bibinfo {author} {\bibfnamefont {B.~M.}\ \bibnamefont
  {Maune}}, \bibinfo {author} {\bibfnamefont {M.~G.}\ \bibnamefont {Borselli}},
  \bibinfo {author} {\bibfnamefont {B.}~\bibnamefont {Huang}}, \bibinfo
  {author} {\bibfnamefont {T.~D.}\ \bibnamefont {Ladd}}, \bibinfo {author}
  {\bibfnamefont {P.~W.}\ \bibnamefont {Deelman}}, \bibinfo {author}
  {\bibfnamefont {K.~S.}\ \bibnamefont {Holabird}}, \bibinfo {author}
  {\bibfnamefont {A.~A.}\ \bibnamefont {Kiselev}}, \bibinfo {author}
  {\bibfnamefont {I.}~\bibnamefont {Alvarado-Rodriguez}}, \bibinfo {author}
  {\bibfnamefont {R.~S.}\ \bibnamefont {Ross}}, \bibinfo {author}
  {\bibfnamefont {A.~E.}\ \bibnamefont {Schmitz}}, \bibinfo {author}
  {\bibfnamefont {M.}~\bibnamefont {Sokolich}}, \bibinfo {author}
  {\bibfnamefont {C.~A.}\ \bibnamefont {Watson}}, \bibinfo {author}
  {\bibfnamefont {M.~F.}\ \bibnamefont {Gyure}}, \ and\ \bibinfo {author}
  {\bibfnamefont {A.~T.}\ \bibnamefont {Hunter}},\ }\href {\doibase
  10.1038/nature10707} {\bibfield  {journal} {\bibinfo  {journal} {Nature}\
  }\textbf {\bibinfo {volume} {481}},\ \bibinfo {pages} {344} (\bibinfo {year}
  {2012})}\BibitemShut {NoStop}%
\bibitem [{\citenamefont {Pla}\ \emph {et~al.}(2012)\citenamefont {Pla},
  \citenamefont {Tan}, \citenamefont {Dehollain}, \citenamefont {Lim},
  \citenamefont {Morton}, \citenamefont {Jamieson}, \citenamefont {Dzurak},\
  and\ \citenamefont {Morello}}]{Pla2012}%
  \BibitemOpen
  \bibfield  {author} {\bibinfo {author} {\bibfnamefont {J.~J.}\ \bibnamefont
  {Pla}}, \bibinfo {author} {\bibfnamefont {K.~Y.}\ \bibnamefont {Tan}},
  \bibinfo {author} {\bibfnamefont {J.~P.}\ \bibnamefont {Dehollain}}, \bibinfo
  {author} {\bibfnamefont {W.~H.}\ \bibnamefont {Lim}}, \bibinfo {author}
  {\bibfnamefont {J.~J.~L.}\ \bibnamefont {Morton}}, \bibinfo {author}
  {\bibfnamefont {D.~N.}\ \bibnamefont {Jamieson}}, \bibinfo {author}
  {\bibfnamefont {A.~S.}\ \bibnamefont {Dzurak}}, \ and\ \bibinfo {author}
  {\bibfnamefont {A.}~\bibnamefont {Morello}},\ }\href {\doibase
  10.1038/nature11449} {\bibfield  {journal} {\bibinfo  {journal} {Nature}\
  }\textbf {\bibinfo {volume} {489}},\ \bibinfo {pages} {541} (\bibinfo {year}
  {2012})}\BibitemShut {NoStop}%
\bibitem [{\citenamefont {Kawakami}\ \emph {et~al.}(2014)\citenamefont
  {Kawakami}, \citenamefont {Scarlino}, \citenamefont {Ward}, \citenamefont
  {Braakman}, \citenamefont {Savage}, \citenamefont {Lagally}, \citenamefont
  {Friesen}, \citenamefont {Coppersmith}, \citenamefont {Eriksson},\ and\
  \citenamefont {Vandersypen}}]{Kawakami2014}%
  \BibitemOpen
  \bibfield  {author} {\bibinfo {author} {\bibfnamefont {E.}~\bibnamefont
  {Kawakami}}, \bibinfo {author} {\bibfnamefont {P.}~\bibnamefont {Scarlino}},
  \bibinfo {author} {\bibfnamefont {D.~R.}\ \bibnamefont {Ward}}, \bibinfo
  {author} {\bibfnamefont {F.~R.}\ \bibnamefont {Braakman}}, \bibinfo {author}
  {\bibfnamefont {D.~E.}\ \bibnamefont {Savage}}, \bibinfo {author}
  {\bibfnamefont {M.~G.}\ \bibnamefont {Lagally}}, \bibinfo {author}
  {\bibfnamefont {M.}~\bibnamefont {Friesen}}, \bibinfo {author} {\bibfnamefont
  {S.~N.}\ \bibnamefont {Coppersmith}}, \bibinfo {author} {\bibfnamefont
  {M.~A.}\ \bibnamefont {Eriksson}}, \ and\ \bibinfo {author} {\bibfnamefont
  {L.~M.~K.}\ \bibnamefont {Vandersypen}},\ }\href {\doibase
  10.1038/nnano.2014.153} {\bibfield  {journal} {\bibinfo  {journal} {Nature
  nanotechnology}\ }\textbf {\bibinfo {volume} {9}},\ \bibinfo {pages} {666}
  (\bibinfo {year} {2014})}\BibitemShut {NoStop}%
\bibitem [{\citenamefont {Veldhorst}\ \emph {et~al.}(2014)\citenamefont
  {Veldhorst}, \citenamefont {Hwang}, \citenamefont {Yang}, \citenamefont
  {Leenstra}, \citenamefont {de~Ronde}, \citenamefont {Dehollain},
  \citenamefont {Muhonen}, \citenamefont {Hudson}, \citenamefont {Itoh},
  \citenamefont {Morello},\ and\ \citenamefont {Dzurak}}]{Veldhorst2014}%
  \BibitemOpen
  \bibfield  {author} {\bibinfo {author} {\bibfnamefont {M.}~\bibnamefont
  {Veldhorst}}, \bibinfo {author} {\bibfnamefont {J.~C.~C.}\ \bibnamefont
  {Hwang}}, \bibinfo {author} {\bibfnamefont {C.~H.}\ \bibnamefont {Yang}},
  \bibinfo {author} {\bibfnamefont {A.~W.}\ \bibnamefont {Leenstra}}, \bibinfo
  {author} {\bibfnamefont {B.}~\bibnamefont {de~Ronde}}, \bibinfo {author}
  {\bibfnamefont {J.~P.}\ \bibnamefont {Dehollain}}, \bibinfo {author}
  {\bibfnamefont {J.~T.}\ \bibnamefont {Muhonen}}, \bibinfo {author}
  {\bibfnamefont {F.~E.}\ \bibnamefont {Hudson}}, \bibinfo {author}
  {\bibfnamefont {K.~M.}\ \bibnamefont {Itoh}}, \bibinfo {author}
  {\bibfnamefont {A.}~\bibnamefont {Morello}}, \ and\ \bibinfo {author}
  {\bibfnamefont {A.~S.}\ \bibnamefont {Dzurak}},\ }\href {\doibase
  10.1038/nnano.2014.216} {\bibfield  {journal} {\bibinfo  {journal} {Nature
  nanotechnology}\ }\textbf {\bibinfo {volume} {9}},\ \bibinfo {pages} {981}
  (\bibinfo {year} {2014})}\BibitemShut {NoStop}%
\bibitem [{\citenamefont {Muhonen}\ \emph {et~al.}(2014)\citenamefont
  {Muhonen}, \citenamefont {Dehollain}, \citenamefont {Laucht}, \citenamefont
  {Hudson}, \citenamefont {Kalra}, \citenamefont {Sekiguchi}, \citenamefont
  {Itoh}, \citenamefont {Jamieson}, \citenamefont {McCallum}, \citenamefont
  {Dzurak},\ and\ \citenamefont {Morello}}]{Muhonen2014}%
  \BibitemOpen
  \bibfield  {author} {\bibinfo {author} {\bibfnamefont {J.~T.}\ \bibnamefont
  {Muhonen}}, \bibinfo {author} {\bibfnamefont {J.~P.}\ \bibnamefont
  {Dehollain}}, \bibinfo {author} {\bibfnamefont {A.}~\bibnamefont {Laucht}},
  \bibinfo {author} {\bibfnamefont {F.~E.}\ \bibnamefont {Hudson}}, \bibinfo
  {author} {\bibfnamefont {R.}~\bibnamefont {Kalra}}, \bibinfo {author}
  {\bibfnamefont {T.}~\bibnamefont {Sekiguchi}}, \bibinfo {author}
  {\bibfnamefont {K.~M.}\ \bibnamefont {Itoh}}, \bibinfo {author}
  {\bibfnamefont {D.~N.}\ \bibnamefont {Jamieson}}, \bibinfo {author}
  {\bibfnamefont {J.~C.}\ \bibnamefont {McCallum}}, \bibinfo {author}
  {\bibfnamefont {A.~S.}\ \bibnamefont {Dzurak}}, \ and\ \bibinfo {author}
  {\bibfnamefont {A.}~\bibnamefont {Morello}},\ }\href {\doibase
  10.1038/nnano.2014.211} {\bibfield  {journal} {\bibinfo  {journal} {Nature
  nanotechnology}\ }\textbf {\bibinfo {volume} {9}},\ \bibinfo {pages} {986}
  (\bibinfo {year} {2014})}\BibitemShut {NoStop}%
\bibitem [{\citenamefont {Veldhorst}\ \emph {et~al.}(2015)\citenamefont
  {Veldhorst}, \citenamefont {Yang}, \citenamefont {Hwang}, \citenamefont
  {Huang}, \citenamefont {Dehollain}, \citenamefont {Muhonen}, \citenamefont
  {Simmons}, \citenamefont {Laucht}, \citenamefont {Hudson}, \citenamefont
  {Itoh}, \citenamefont {Morello},\ and\ \citenamefont
  {Dzurak}}]{Veldhorst2015}%
  \BibitemOpen
  \bibfield  {author} {\bibinfo {author} {\bibfnamefont {M.}~\bibnamefont
  {Veldhorst}}, \bibinfo {author} {\bibfnamefont {C.~H.}\ \bibnamefont {Yang}},
  \bibinfo {author} {\bibfnamefont {J.~C.~C.}\ \bibnamefont {Hwang}}, \bibinfo
  {author} {\bibfnamefont {W.}~\bibnamefont {Huang}}, \bibinfo {author}
  {\bibfnamefont {J.~P.}\ \bibnamefont {Dehollain}}, \bibinfo {author}
  {\bibfnamefont {J.~T.}\ \bibnamefont {Muhonen}}, \bibinfo {author}
  {\bibfnamefont {S.}~\bibnamefont {Simmons}}, \bibinfo {author} {\bibfnamefont
  {a.}~\bibnamefont {Laucht}}, \bibinfo {author} {\bibfnamefont {F.~E.}\
  \bibnamefont {Hudson}}, \bibinfo {author} {\bibfnamefont {K.~M.}\
  \bibnamefont {Itoh}}, \bibinfo {author} {\bibfnamefont {a.}~\bibnamefont
  {Morello}}, \ and\ \bibinfo {author} {\bibfnamefont {a.~S.}\ \bibnamefont
  {Dzurak}},\ }\href {\doibase 10.1038/nature15263} {\bibfield  {journal}
  {\bibinfo  {journal} {Nature}\ ,\ \bibinfo {pages} {0}} (\bibinfo {year}
  {2015})}\BibitemShut {NoStop}%
\bibitem [{\citenamefont {Kane}(1998)}]{Kane1998}%
  \BibitemOpen
  \bibfield  {author} {\bibinfo {author} {\bibfnamefont {B.~E.}\ \bibnamefont
  {Kane}},\ }\href@noop {} {\bibfield  {journal} {\bibinfo  {journal} {Nature}\
  }\textbf {\bibinfo {volume} {393}},\ \bibinfo {pages} {133} (\bibinfo {year}
  {1998})}\BibitemShut {NoStop}%
\bibitem [{\citenamefont {Loss}\ and\ \citenamefont
  {Divincenzo}(1998)}]{Loss1998}%
  \BibitemOpen
  \bibfield  {author} {\bibinfo {author} {\bibfnamefont {D.}~\bibnamefont
  {Loss}}\ and\ \bibinfo {author} {\bibfnamefont {D.~P.}\ \bibnamefont
  {Divincenzo}},\ }\href@noop {} {\bibfield  {journal} {\bibinfo  {journal}
  {Phys. Rev. A}\ }\textbf {\bibinfo {volume} {57}},\ \bibinfo {pages} {120}
  (\bibinfo {year} {1998})}\BibitemShut {NoStop}%
\bibitem [{\citenamefont {Hill}\ \emph {et~al.}(2015)\citenamefont {Hill},
  \citenamefont {Peretz}, \citenamefont {Hile}, \citenamefont {House},
  \citenamefont {Fuechsle}, \citenamefont {Rogge}, \citenamefont {Simmons},\
  and\ \citenamefont {Hollenberg}}]{Hill2015}%
  \BibitemOpen
  \bibfield  {author} {\bibinfo {author} {\bibfnamefont {C.~D.}\ \bibnamefont
  {Hill}}, \bibinfo {author} {\bibfnamefont {E.}~\bibnamefont {Peretz}},
  \bibinfo {author} {\bibfnamefont {S.~J.}\ \bibnamefont {Hile}}, \bibinfo
  {author} {\bibfnamefont {M.~G.}\ \bibnamefont {House}}, \bibinfo {author}
  {\bibfnamefont {M.}~\bibnamefont {Fuechsle}}, \bibinfo {author}
  {\bibfnamefont {S.}~\bibnamefont {Rogge}}, \bibinfo {author} {\bibfnamefont
  {M.~Y.}\ \bibnamefont {Simmons}}, \ and\ \bibinfo {author} {\bibfnamefont
  {L.~C.~L.}\ \bibnamefont {Hollenberg}},\ }\href@noop {} {\bibfield  {journal}
  {\bibinfo  {journal} {Science Advances}\ }\textbf {\bibinfo {volume}
  {e1500707}},\ \bibinfo {pages} {1} (\bibinfo {year} {2015})}\BibitemShut
  {NoStop}%
\bibitem [{\citenamefont {Pica}\ \emph {et~al.}(2016)\citenamefont {Pica},
  \citenamefont {Lovett}, \citenamefont {Bhatt}, \citenamefont {Schenkel},\
  and\ \citenamefont {Lyon}}]{Pica2016}%
  \BibitemOpen
  \bibfield  {author} {\bibinfo {author} {\bibfnamefont {G.}~\bibnamefont
  {Pica}}, \bibinfo {author} {\bibfnamefont {B.~W.}\ \bibnamefont {Lovett}},
  \bibinfo {author} {\bibfnamefont {R.~N.}\ \bibnamefont {Bhatt}}, \bibinfo
  {author} {\bibfnamefont {T.}~\bibnamefont {Schenkel}}, \ and\ \bibinfo
  {author} {\bibfnamefont {S.~a.}\ \bibnamefont {Lyon}},\ }\href {\doibase
  10.1103/PhysRevB.93.035306} {\bibfield  {journal} {\bibinfo  {journal}
  {Physical Review B}\ }\textbf {\bibinfo {volume} {93}},\ \bibinfo {pages}
  {035306} (\bibinfo {year} {2016})}\BibitemShut {NoStop}%
\bibitem [{\citenamefont {Levy}\ \emph {et~al.}(2011)\citenamefont {Levy},
  \citenamefont {Carroll}, \citenamefont {Ganti}, \citenamefont {Phillips},
  \citenamefont {Landahl}, \citenamefont {Gurrieri}, \citenamefont {Carr},
  \citenamefont {Stalford},\ and\ \citenamefont {Nielsen}}]{Levy2011}%
  \BibitemOpen
  \bibfield  {author} {\bibinfo {author} {\bibfnamefont {J.~E.}\ \bibnamefont
  {Levy}}, \bibinfo {author} {\bibfnamefont {M.~S.}\ \bibnamefont {Carroll}},
  \bibinfo {author} {\bibfnamefont {A.}~\bibnamefont {Ganti}}, \bibinfo
  {author} {\bibfnamefont {C.~A.}\ \bibnamefont {Phillips}}, \bibinfo {author}
  {\bibfnamefont {A.~J.}\ \bibnamefont {Landahl}}, \bibinfo {author}
  {\bibfnamefont {T.~M.}\ \bibnamefont {Gurrieri}}, \bibinfo {author}
  {\bibfnamefont {R.~D.}\ \bibnamefont {Carr}}, \bibinfo {author}
  {\bibfnamefont {H.~L.}\ \bibnamefont {Stalford}}, \ and\ \bibinfo {author}
  {\bibfnamefont {E.}~\bibnamefont {Nielsen}},\ }\href {\doibase
  10.1088/1367-2630/13/8/083021} {\bibfield  {journal} {\bibinfo  {journal}
  {New Journal of Physics}\ }\textbf {\bibinfo {volume} {13}},\ \bibinfo
  {pages} {083021} (\bibinfo {year} {2011})}\BibitemShut {NoStop}%
\bibitem [{\citenamefont {Barraud}\ \emph {et~al.}(2012)\citenamefont
  {Barraud}, \citenamefont {Coquand}, \citenamefont {Cass\'{e}}, \citenamefont
  {Koyama}, \citenamefont {Hartmann}, \citenamefont {Comboroure}, \citenamefont
  {Vizioz}, \citenamefont {Aussenac}, \citenamefont {Faynot},\ and\
  \citenamefont {Poiroux}}]{Barraud2012}%
  \BibitemOpen
  \bibfield  {author} {\bibinfo {author} {\bibfnamefont {S.}~\bibnamefont
  {Barraud}}, \bibinfo {author} {\bibfnamefont {R.}~\bibnamefont {Coquand}},
  \bibinfo {author} {\bibfnamefont {M.}~\bibnamefont {Cass\'{e}}}, \bibinfo
  {author} {\bibfnamefont {M.}~\bibnamefont {Koyama}}, \bibinfo {author}
  {\bibfnamefont {J.}~\bibnamefont {Hartmann}}, \bibinfo {author}
  {\bibfnamefont {C.}~\bibnamefont {Comboroure}}, \bibinfo {author}
  {\bibfnamefont {C.}~\bibnamefont {Vizioz}}, \bibinfo {author} {\bibfnamefont
  {F.}~\bibnamefont {Aussenac}}, \bibinfo {author} {\bibfnamefont
  {O.}~\bibnamefont {Faynot}}, \ and\ \bibinfo {author} {\bibfnamefont
  {T.}~\bibnamefont {Poiroux}},\ }\href@noop {} {\bibfield  {journal} {\bibinfo
   {journal} {IEEE Electron Device Letter}\ }\textbf {\bibinfo {volume} {33}},\
  \bibinfo {pages} {1526} (\bibinfo {year} {2012})}\BibitemShut {NoStop}%
\bibitem [{\citenamefont {Petta}\ \emph {et~al.}(2005)\citenamefont {Petta},
  \citenamefont {Johnson}, \citenamefont {Taylor}, \citenamefont {Laird},
  \citenamefont {Yacoby}, \citenamefont {Lukin}, \citenamefont {Marcus},
  \citenamefont {Hanson},\ and\ \citenamefont {Gossard}}]{Petta2005}%
  \BibitemOpen
  \bibfield  {author} {\bibinfo {author} {\bibfnamefont {J.~R.}\ \bibnamefont
  {Petta}}, \bibinfo {author} {\bibfnamefont {A.~C.}\ \bibnamefont {Johnson}},
  \bibinfo {author} {\bibfnamefont {J.~M.}\ \bibnamefont {Taylor}}, \bibinfo
  {author} {\bibfnamefont {E.~A.}\ \bibnamefont {Laird}}, \bibinfo {author}
  {\bibfnamefont {A.}~\bibnamefont {Yacoby}}, \bibinfo {author} {\bibfnamefont
  {M.~D.}\ \bibnamefont {Lukin}}, \bibinfo {author} {\bibfnamefont {C.~M.}\
  \bibnamefont {Marcus}}, \bibinfo {author} {\bibfnamefont {M.~P.}\
  \bibnamefont {Hanson}}, \ and\ \bibinfo {author} {\bibfnamefont {A.~C.}\
  \bibnamefont {Gossard}},\ }\href {\doibase 10.1126/science.1116955}
  {\bibfield  {journal} {\bibinfo  {journal} {Science (New York, N.Y.)}\
  }\textbf {\bibinfo {volume} {309}},\ \bibinfo {pages} {2180} (\bibinfo {year}
  {2005})}\BibitemShut {NoStop}%
\bibitem [{\citenamefont {Koppens}\ \emph {et~al.}(2006)\citenamefont
  {Koppens}, \citenamefont {Buizert}, \citenamefont {Tielrooij}, \citenamefont
  {Vink}, \citenamefont {Nowack}, \citenamefont {Meunier}, \citenamefont
  {Kouwenhoven},\ and\ \citenamefont {Vandersypen}}]{Koppens2006}%
  \BibitemOpen
  \bibfield  {author} {\bibinfo {author} {\bibfnamefont {F.~H.~L.}\
  \bibnamefont {Koppens}}, \bibinfo {author} {\bibfnamefont {C.}~\bibnamefont
  {Buizert}}, \bibinfo {author} {\bibfnamefont {K.~J.}\ \bibnamefont
  {Tielrooij}}, \bibinfo {author} {\bibfnamefont {I.~T.}\ \bibnamefont {Vink}},
  \bibinfo {author} {\bibfnamefont {K.~C.}\ \bibnamefont {Nowack}}, \bibinfo
  {author} {\bibfnamefont {T.}~\bibnamefont {Meunier}}, \bibinfo {author}
  {\bibfnamefont {L.~P.}\ \bibnamefont {Kouwenhoven}}, \ and\ \bibinfo {author}
  {\bibfnamefont {L.~M.~K.}\ \bibnamefont {Vandersypen}},\ }\href {\doibase
  10.1038/nature05065} {\bibfield  {journal} {\bibinfo  {journal} {Nature}\
  }\textbf {\bibinfo {volume} {442}},\ \bibinfo {pages} {766} (\bibinfo {year}
  {2006})}\BibitemShut {NoStop}%
\bibitem [{\citenamefont {Hanson}\ \emph {et~al.}(2007)\citenamefont {Hanson},
  \citenamefont {Kouwenhoven},\ and\ \citenamefont {Petta}}]{Hanson2006}%
  \BibitemOpen
  \bibfield  {author} {\bibinfo {author} {\bibfnamefont {R.}~\bibnamefont
  {Hanson}}, \bibinfo {author} {\bibfnamefont {L.}~\bibnamefont {Kouwenhoven}},
  \ and\ \bibinfo {author} {\bibfnamefont {J.}~\bibnamefont {Petta}},\ }\href
  {http://scitation.aip.org/getpdf/servlet/GetPDFServlet?filetype=pdf\&id=RMPHAT000079000004001217000001\&idtype=cvips\&prog=normal
  http://journals.aps.org/rmp/abstract/10.1103/RevModPhys.79.1217} {\bibfield
  {journal} {\bibinfo  {journal} {Reviews of Modern Physics}\ }\textbf
  {\bibinfo {volume} {79}} (\bibinfo {year} {2007})}\BibitemShut {NoStop}%
\bibitem [{\citenamefont {Koppens}\ \emph {et~al.}(2008)\citenamefont
  {Koppens}, \citenamefont {Nowack},\ and\ \citenamefont
  {Vandersypen}}]{Koppens2008}%
  \BibitemOpen
  \bibfield  {author} {\bibinfo {author} {\bibfnamefont {F.~H.~L.}\
  \bibnamefont {Koppens}}, \bibinfo {author} {\bibfnamefont {K.~C.}\
  \bibnamefont {Nowack}}, \ and\ \bibinfo {author} {\bibfnamefont {L.~M.~K.}\
  \bibnamefont {Vandersypen}},\ }\href {\doibase
  10.1103/PhysRevLett.100.236802} {\bibfield  {journal} {\bibinfo  {journal}
  {Physical Review Letters}\ }\textbf {\bibinfo {volume} {100}},\ \bibinfo
  {pages} {236802} (\bibinfo {year} {2008})}\BibitemShut {NoStop}%
\bibitem [{\citenamefont {Bluhm}\ \emph {et~al.}(2010)\citenamefont {Bluhm},
  \citenamefont {Foletti}, \citenamefont {Neder}, \citenamefont {Rudner},
  \citenamefont {Mahalu}, \citenamefont {Umansky},\ and\ \citenamefont
  {Yacoby}}]{Bluhm2010_2}%
  \BibitemOpen
  \bibfield  {author} {\bibinfo {author} {\bibfnamefont {H.}~\bibnamefont
  {Bluhm}}, \bibinfo {author} {\bibfnamefont {S.}~\bibnamefont {Foletti}},
  \bibinfo {author} {\bibfnamefont {I.}~\bibnamefont {Neder}}, \bibinfo
  {author} {\bibfnamefont {M.}~\bibnamefont {Rudner}}, \bibinfo {author}
  {\bibfnamefont {D.}~\bibnamefont {Mahalu}}, \bibinfo {author} {\bibfnamefont
  {V.}~\bibnamefont {Umansky}}, \ and\ \bibinfo {author} {\bibfnamefont
  {A.}~\bibnamefont {Yacoby}},\ }\href {\doibase 10.1038/nphys1856} {\bibfield
  {journal} {\bibinfo  {journal} {Nature Physics}\ }\textbf {\bibinfo {volume}
  {7}},\ \bibinfo {pages} {109} (\bibinfo {year} {2010})}\BibitemShut {NoStop}%
\bibitem [{\citenamefont {Lange}\ \emph {et~al.}(2010)\citenamefont {Lange},
  \citenamefont {Wang}, \citenamefont {Dobrovitski},\ and\ \citenamefont
  {Hanson}}]{Lange2010}%
  \BibitemOpen
  \bibfield  {author} {\bibinfo {author} {\bibfnamefont {G.~D.}\ \bibnamefont
  {Lange}}, \bibinfo {author} {\bibfnamefont {Z.~H.}\ \bibnamefont {Wang}},
  \bibinfo {author} {\bibfnamefont {V.~V.}\ \bibnamefont {Dobrovitski}}, \ and\
  \bibinfo {author} {\bibfnamefont {R.}~\bibnamefont {Hanson}},\ }\href@noop {}
  {\bibfield  {journal} {\bibinfo  {journal} {Science}\ }\textbf {\bibinfo
  {volume} {330}},\ \bibinfo {pages} {60} (\bibinfo {year} {2010})}\BibitemShut
  {NoStop}%
\bibitem [{\citenamefont {Malinowski}\ \emph {et~al.}(2016)\citenamefont
  {Malinowski}, \citenamefont {Martins}, \citenamefont {Nissen}, \citenamefont
  {Barnes}, \citenamefont {Rudner}, \citenamefont {Fallahi}, \citenamefont
  {Gardner}, \citenamefont {Manfra}, \citenamefont {Marcus},\ and\
  \citenamefont {Kuemmeth}}]{Malinowski2016}%
  \BibitemOpen
  \bibfield  {author} {\bibinfo {author} {\bibfnamefont {F.~K.}\ \bibnamefont
  {Malinowski}}, \bibinfo {author} {\bibfnamefont {F.}~\bibnamefont {Martins}},
  \bibinfo {author} {\bibfnamefont {P.~D.}\ \bibnamefont {Nissen}}, \bibinfo
  {author} {\bibfnamefont {E.}~\bibnamefont {Barnes}}, \bibinfo {author}
  {\bibfnamefont {M.~S.}\ \bibnamefont {Rudner}}, \bibinfo {author}
  {\bibfnamefont {S.}~\bibnamefont {Fallahi}}, \bibinfo {author} {\bibfnamefont
  {G.~C.}\ \bibnamefont {Gardner}}, \bibinfo {author} {\bibfnamefont {M.~J.}\
  \bibnamefont {Manfra}}, \bibinfo {author} {\bibfnamefont {C.~M.}\
  \bibnamefont {Marcus}}, \ and\ \bibinfo {author} {\bibfnamefont
  {F.}~\bibnamefont {Kuemmeth}},\ }\href@noop {} {\bibfield  {journal}
  {\bibinfo  {journal} {arXiv:1601.0667}\ } (\bibinfo {year}
  {2016})}\BibitemShut {NoStop}%
\bibitem [{\citenamefont {Wu}\ \emph {et~al.}(2014)\citenamefont {Wu},
  \citenamefont {Ward}, \citenamefont {Prance}, \citenamefont {Kim},
  \citenamefont {Gamble}, \citenamefont {Mohr}, \citenamefont {Shi},
  \citenamefont {Savage}, \citenamefont {Lagally}, \citenamefont {Friesen},
  \citenamefont {Coppersmith},\ and\ \citenamefont {Eriksson}}]{Wu2014}%
  \BibitemOpen
  \bibfield  {author} {\bibinfo {author} {\bibfnamefont {X.}~\bibnamefont
  {Wu}}, \bibinfo {author} {\bibfnamefont {D.~R.}\ \bibnamefont {Ward}},
  \bibinfo {author} {\bibfnamefont {J.~R.}\ \bibnamefont {Prance}}, \bibinfo
  {author} {\bibfnamefont {D.}~\bibnamefont {Kim}}, \bibinfo {author}
  {\bibfnamefont {J.~K.}\ \bibnamefont {Gamble}}, \bibinfo {author}
  {\bibfnamefont {R.~T.}\ \bibnamefont {Mohr}}, \bibinfo {author}
  {\bibfnamefont {Z.}~\bibnamefont {Shi}}, \bibinfo {author} {\bibfnamefont
  {D.~E.}\ \bibnamefont {Savage}}, \bibinfo {author} {\bibfnamefont {M.~G.}\
  \bibnamefont {Lagally}}, \bibinfo {author} {\bibfnamefont {M.}~\bibnamefont
  {Friesen}}, \bibinfo {author} {\bibfnamefont {S.~N.}\ \bibnamefont
  {Coppersmith}}, \ and\ \bibinfo {author} {\bibfnamefont {M.~a.}\ \bibnamefont
  {Eriksson}},\ }\href {\doibase 10.1073/pnas.1412230111} {\bibfield  {journal}
  {\bibinfo  {journal} {Proceedings of the National Academy of Sciences of the
  United States of America}\ }\textbf {\bibinfo {volume} {111}},\ \bibinfo
  {pages} {11938} (\bibinfo {year} {2014})}\BibitemShut {NoStop}%
\bibitem [{\citenamefont {Bravyi}\ and\ \citenamefont {Kitaev}()}]{Bravyi2008}%
  \BibitemOpen
  \bibfield  {author} {\bibinfo {author} {\bibfnamefont {S.~B.}\ \bibnamefont
  {Bravyi}}\ and\ \bibinfo {author} {\bibfnamefont {A.~Y.}\ \bibnamefont
  {Kitaev}},\ }\href@noop {} {\bibinfo  {journal} {arXiv:quant-ph/9811052v1}\
  }\BibitemShut {NoStop}%
\bibitem [{\citenamefont {Dennis}\ \emph {et~al.}(2002)\citenamefont {Dennis},
  \citenamefont {Kitaev}, \citenamefont {Landahl},\ and\ \citenamefont
  {Preskill}}]{Dennis2002}%
  \BibitemOpen
\bibfield  {journal} {  }\bibfield  {author} {\bibinfo {author} {\bibfnamefont
  {E.}~\bibnamefont {Dennis}}, \bibinfo {author} {\bibfnamefont
  {A.}~\bibnamefont {Kitaev}}, \bibinfo {author} {\bibfnamefont
  {A.}~\bibnamefont {Landahl}}, \ and\ \bibinfo {author} {\bibfnamefont
  {J.}~\bibnamefont {Preskill}},\ }\href@noop {} {\bibfield  {journal}
  {\bibinfo  {journal} {Journal of Mathematical Physics}\ }\textbf {\bibinfo
  {volume} {43}},\ \bibinfo {pages} {4452} (\bibinfo {year}
  {2002})}\BibitemShut {NoStop}%
\bibitem [{\citenamefont {Voisin}\ \emph {et~al.}(2016)\citenamefont {Voisin},
  \citenamefont {Maurand}, \citenamefont {Barraud}, \citenamefont {Vinet},
  \citenamefont {Jehl}, \citenamefont {Sanquer}, \citenamefont {Renard},\ and\
  \citenamefont {{De Franceschi}}}]{Voisin2016}%
  \BibitemOpen
  \bibfield  {author} {\bibinfo {author} {\bibfnamefont {B.}~\bibnamefont
  {Voisin}}, \bibinfo {author} {\bibfnamefont {R.}~\bibnamefont {Maurand}},
  \bibinfo {author} {\bibfnamefont {S.}~\bibnamefont {Barraud}}, \bibinfo
  {author} {\bibfnamefont {M.}~\bibnamefont {Vinet}}, \bibinfo {author}
  {\bibfnamefont {X.}~\bibnamefont {Jehl}}, \bibinfo {author} {\bibfnamefont
  {M.}~\bibnamefont {Sanquer}}, \bibinfo {author} {\bibfnamefont
  {J.}~\bibnamefont {Renard}}, \ and\ \bibinfo {author} {\bibfnamefont
  {S.}~\bibnamefont {{De Franceschi}}},\ }\href {\doibase
  10.1021/acs.nanolett.5b02920} {\bibfield  {journal} {\bibinfo  {journal}
  {Nano letters}\ }\textbf {\bibinfo {volume} {16}},\ \bibinfo {pages} {88}
  (\bibinfo {year} {2016})}\BibitemShut {NoStop}%
\bibitem [{\citenamefont {Ono}\ \emph {et~al.}(2002)\citenamefont {Ono},
  \citenamefont {Austing}, \citenamefont {Tokura},\ and\ \citenamefont
  {Tarucha}}]{Ono2002}%
  \BibitemOpen
  \bibfield  {author} {\bibinfo {author} {\bibfnamefont {K.}~\bibnamefont
  {Ono}}, \bibinfo {author} {\bibfnamefont {D.}~\bibnamefont {Austing}},
  \bibinfo {author} {\bibfnamefont {Y.}~\bibnamefont {Tokura}}, \ and\ \bibinfo
  {author} {\bibfnamefont {S.}~\bibnamefont {Tarucha}},\ }\href
  {http://science.sciencemag.org/content/297/5585/1313.short} {\bibfield
  {journal} {\bibinfo  {journal} {Science}\ }\textbf {\bibinfo {volume}
  {297}},\ \bibinfo {pages} {1313} (\bibinfo {year} {2002})}\BibitemShut
  {NoStop}%
\bibitem [{\citenamefont {Danon}\ and\ \citenamefont
  {Nazarov}(2009)}]{Danon2009}%
  \BibitemOpen
  \bibfield  {author} {\bibinfo {author} {\bibfnamefont {J.}~\bibnamefont
  {Danon}}\ and\ \bibinfo {author} {\bibfnamefont {Y.~V.}\ \bibnamefont
  {Nazarov}},\ }\href {\doibase 10.1103/PhysRevB.80.041301} {\bibfield
  {journal} {\bibinfo  {journal} {Physical Review B}\ }\textbf {\bibinfo
  {volume} {80}},\ \bibinfo {pages} {041301} (\bibinfo {year}
  {2009})}\BibitemShut {NoStop}%
\bibitem [{\citenamefont {Nadj-Perge}\ \emph {et~al.}(2010)\citenamefont
  {Nadj-Perge}, \citenamefont {Frolov}, \citenamefont {van Tilburg},
  \citenamefont {Danon}, \citenamefont {Nazarov}, \citenamefont {Algra},
  \citenamefont {Bakkers},\ and\ \citenamefont {Kouwenhoven}}]{Nadj-Perge2010}%
  \BibitemOpen
  \bibfield  {author} {\bibinfo {author} {\bibfnamefont {S.}~\bibnamefont
  {Nadj-Perge}}, \bibinfo {author} {\bibfnamefont {S.~M.}\ \bibnamefont
  {Frolov}}, \bibinfo {author} {\bibfnamefont {J.~W.~W.}\ \bibnamefont {van
  Tilburg}}, \bibinfo {author} {\bibfnamefont {J.}~\bibnamefont {Danon}},
  \bibinfo {author} {\bibfnamefont {Y.~V.}\ \bibnamefont {Nazarov}}, \bibinfo
  {author} {\bibfnamefont {R.}~\bibnamefont {Algra}}, \bibinfo {author}
  {\bibfnamefont {E.~P. A.~M.}\ \bibnamefont {Bakkers}}, \ and\ \bibinfo
  {author} {\bibfnamefont {L.~P.}\ \bibnamefont {Kouwenhoven}},\ }\href
  {\doibase 10.1103/PhysRevB.81.201305} {\bibfield  {journal} {\bibinfo
  {journal} {Physical Review B}\ }\textbf {\bibinfo {volume} {81}},\ \bibinfo
  {pages} {201305} (\bibinfo {year} {2010})}\BibitemShut {NoStop}%
\bibitem [{\citenamefont {Li}\ \emph {et~al.}(2015)\citenamefont {Li},
  \citenamefont {Hudson}, \citenamefont {Dzurak},\ and\ \citenamefont
  {Hamilton}}]{Li2015}%
  \BibitemOpen
  \bibfield  {author} {\bibinfo {author} {\bibfnamefont {R.}~\bibnamefont
  {Li}}, \bibinfo {author} {\bibfnamefont {F.~E.}\ \bibnamefont {Hudson}},
  \bibinfo {author} {\bibfnamefont {A.~S.}\ \bibnamefont {Dzurak}}, \ and\
  \bibinfo {author} {\bibfnamefont {A.~R.}\ \bibnamefont {Hamilton}},\ }\href
  {\doibase 10.1021/acs.nanolett.5b02561} {\bibfield  {journal} {\bibinfo
  {journal} {Nano letters}\ }\textbf {\bibinfo {volume} {15}},\ \bibinfo
  {pages} {7314} (\bibinfo {year} {2015})}\BibitemShut {NoStop}%
\bibitem [{\citenamefont {Koppens}\ \emph {et~al.}(2007)\citenamefont
  {Koppens}, \citenamefont {Klauser}, \citenamefont {Coish}, \citenamefont
  {Nowack}, \citenamefont {Kouwenhoven}, \citenamefont {Loss},\ and\
  \citenamefont {Vandersypen}}]{Koppens2007}%
  \BibitemOpen
  \bibfield  {author} {\bibinfo {author} {\bibfnamefont {F.~H.~L.}\
  \bibnamefont {Koppens}}, \bibinfo {author} {\bibfnamefont {D.}~\bibnamefont
  {Klauser}}, \bibinfo {author} {\bibfnamefont {W.~a.}\ \bibnamefont {Coish}},
  \bibinfo {author} {\bibfnamefont {K.~C.}\ \bibnamefont {Nowack}}, \bibinfo
  {author} {\bibfnamefont {L.~P.}\ \bibnamefont {Kouwenhoven}}, \bibinfo
  {author} {\bibfnamefont {D.}~\bibnamefont {Loss}}, \ and\ \bibinfo {author}
  {\bibfnamefont {L.~M.~K.}\ \bibnamefont {Vandersypen}},\ }\href {\doibase
  10.1103/PhysRevLett.99.106803} {\bibfield  {journal} {\bibinfo  {journal}
  {Physical review letters}\ }\textbf {\bibinfo {volume} {99}},\ \bibinfo
  {pages} {106803} (\bibinfo {year} {2007})}\BibitemShut {NoStop}%
\bibitem [{\citenamefont {van~den Berg}\ \emph {et~al.}(2013)\citenamefont
  {van~den Berg}, \citenamefont {Nadj-Perge}, \citenamefont {Pribiag},
  \citenamefont {Plissard}, \citenamefont {Bakkers}, \citenamefont {Frolov},\
  and\ \citenamefont {Kouwenhoven}}]{VandenBerg2013}%
  \BibitemOpen
  \bibfield  {author} {\bibinfo {author} {\bibfnamefont {J.~W.~G.}\
  \bibnamefont {van~den Berg}}, \bibinfo {author} {\bibfnamefont
  {S.}~\bibnamefont {Nadj-Perge}}, \bibinfo {author} {\bibfnamefont {V.~S.}\
  \bibnamefont {Pribiag}}, \bibinfo {author} {\bibfnamefont {S.~R.}\
  \bibnamefont {Plissard}}, \bibinfo {author} {\bibfnamefont {E.~P. A.~M.}\
  \bibnamefont {Bakkers}}, \bibinfo {author} {\bibfnamefont {S.~M.}\
  \bibnamefont {Frolov}}, \ and\ \bibinfo {author} {\bibfnamefont {L.~P.}\
  \bibnamefont {Kouwenhoven}},\ }\href {\doibase
  10.1103/PhysRevLett.110.066806} {\bibfield  {journal} {\bibinfo  {journal}
  {Physical Review Letters}\ }\textbf {\bibinfo {volume} {110}},\ \bibinfo
  {pages} {066806} (\bibinfo {year} {2013})}\BibitemShut {NoStop}%
\bibitem [{\citenamefont {Testelin}\ \emph {et~al.}(2009)\citenamefont
  {Testelin}, \citenamefont {Bernardot}, \citenamefont {Eble},\ and\
  \citenamefont {Chamarro}}]{Testelin2009}%
  \BibitemOpen
  \bibfield  {author} {\bibinfo {author} {\bibfnamefont {C.}~\bibnamefont
  {Testelin}}, \bibinfo {author} {\bibfnamefont {F.}~\bibnamefont {Bernardot}},
  \bibinfo {author} {\bibfnamefont {B.}~\bibnamefont {Eble}}, \ and\ \bibinfo
  {author} {\bibfnamefont {M.}~\bibnamefont {Chamarro}},\ }\href {\doibase
  10.1103/PhysRevB.79.195440} {\bibfield  {journal} {\bibinfo  {journal}
  {Physical Review B}\ }\textbf {\bibinfo {volume} {79}},\ \bibinfo {pages}
  {195440} (\bibinfo {year} {2009})}\BibitemShut {NoStop}%
\end{thebibliography}
\end{document}